\documentclass[conference]{IEEEtran}
\IEEEoverridecommandlockouts
\usepackage{cite}
\usepackage{amsmath,amssymb,amsfonts}
\usepackage{algorithmic}
\usepackage{graphicx}
\usepackage{textcomp}
\usepackage[dvipsnames]{xcolor}
\def\BibTeX{{\rm B\kern-.05em{\sc i\kern-.025em b}\kern-.08em
    T\kern-.1667em\lower.7ex\hbox{E}\kern-.125emX}}

\usepackage{url, hyperref}

\newcommand{\R}{\mathbb{R}}

\DeclareFixedFont{\ttb}{T1}{txtt}{bx}{n}{8} 
\DeclareFixedFont{\ttm}{T1}{txtt}{m}{n}{8}  

\usepackage{color}
\definecolor{deepblue}{rgb}{0,0,0.5}
\definecolor{deepred}{rgb}{0.6,0,0}
\definecolor{deepgreen}{rgb}{0,0.5,0}

\usepackage{listings}
\lstdefinestyle{python-style}{
    language=Python,
    basicstyle=\ttm, 
    commentstyle=\ttm\color{deepgreen},
    morekeywords={def}, 
    keywordstyle=\ttb\color{deepblue},
    emph={feature_map, unsqueeze, @, /, cumsum, squeeze}, 
    emphstyle=\ttb\color{deepred}, 
    stringstyle=\color{deepgreen},
    showstringspaces=false,
    numbers=left, 
    breaklines=true,
    showtabs=false,
    tabsize=2,
    xleftmargin=1.4\parindent,
    aboveskip=5pt, 
    belowcaptionskip=5pt,
}

\lstdefinestyle{C-style}{
	belowcaptionskip=1\baselineskip,
	breaklines=true,
	xleftmargin=\parindent,
	language=C,
	showstringspaces=false,
	basicstyle=\footnotesize\ttfamily,
	keywordstyle=\bfseries\color{RoyalBlue},
	commentstyle=\color{ForestGreen},
	stringstyle=\color{ForestGreen},
	escapeinside={(*@}{@*)},
	morekeywords={tensor, float64, int5},
	numbers=left,
        showtabs=false,                  
        tabsize=2,
	morekeywords={[2]{__local__, get_index_space_offset, get_index_space_size, v_f32_ld_tnsr_b, v_exp_cephes_fast_f32, v_f32_reduce_add, v_reciprocal_f32, v_f32_st_tnsr, v_broadcast_element_f32, v_f32_mul_b}},
	keywordstyle={[2]{\bfseries\color{YellowOrange}}},
}  

\usepackage{pifont}

\usepackage[normalem]{ulem}

\usepackage[ruled, linesnumbered, norelsize]{algorithm2e}
\usepackage{array}
\usepackage{booktabs, multirow}

\usepackage{tikz}\usetikzlibrary{positioning,calc}
\newcommand*\Circled[1]{
	\tikz[baseline=(char.base)]{\node[shape=circle, draw=none, fill=gray!40, thick, inner sep=0.6pt] (char) {
			\textcolor{black}{\sffamily#1}}; }}

\usepackage{fancyhdr}
\begin{document}

\title{GFormer: Accelerating Large Language Models with Optimized Transformers on Gaudi Processors}

\author{\IEEEauthorblockN{Chengming Zhang}
\IEEEauthorblockA{
\textit{University of Houston} \\
czhang59@cougarnet.uh.edu}
\and
\IEEEauthorblockN{Xinheng Ding}
\IEEEauthorblockA{
\textit{Indiana University}\\
xinhding@iu.edu}
\and
\IEEEauthorblockN{Baixi Sun}
\IEEEauthorblockA{
\textit{Indiana University}\\
sunbaix@iu.edu}
\and
\IEEEauthorblockN{Xiaodong Yu}
\IEEEauthorblockA{
\textit{Stevens Institute of Technology}\\
xyu38@stevens.edu}
\and
\IEEEauthorblockN{Weijian Zheng}
\IEEEauthorblockA{
\textit{Argonne National Laboratory}\\
wzheng@anl.gov}
\and
\IEEEauthorblockN{Zhen Xie}
\IEEEauthorblockA{
\textit{Binghamton University}\\
zxie3@binghamton.edu}
\and
\IEEEauthorblockN{Dingwen Tao}
\IEEEauthorblockA{
\textit{Indiana University}\\
ditao@iu.edu}
}

\maketitle

\begin{abstract}

Heterogeneous hardware like Gaudi processor has been developed to enhance computations, especially matrix operations for Transformer-based large language models (LLMs) for generative AI tasks. However, our analysis indicates that Transformers are not fully optimized on such emerging hardware, primarily due to inadequate optimizations in non-matrix computational kernels like Softmax and in heterogeneous resource utilization, particularly when processing long sequences. To address these issues, we propose an integrated approach (called GFormer) that merges sparse and linear attention mechanisms. GFormer aims to maximize the computational capabilities of the Gaudi processor's Matrix Multiplication Engine (MME) and Tensor Processing Cores (TPC) without compromising model quality. GFormer includes a windowed self-attention kernel and an efficient outer product kernel for causal linear attention, aiming to optimize LLM inference on Gaudi processors. Evaluation shows that GFormer significantly improves efficiency and model performance across various tasks on the Gaudi processor and outperforms state-of-the-art GPUs.
\end{abstract}


\section{Introduction}
\label{sec:introduction}
Transformers \cite{vaswani2017attention} have revolutionized the field of natural language processing (NLP) and beyond, becoming the backbone of numerous state-of-the-art (SOTA) machine learning applications across machine translation \cite{devlin2018bert}, question answering, and computer vision \cite{dosovitskiy2020image}. Despite Transformers' broad applicability and potential to catalyze advancements across various fields, the computation efficiency of Transformers presents a significant challenge that hampers their broader adoption and scalability. At the heart of this issue is the self-attention mechanism requires quadratic memory and time complexity $O(N^2)$ for processing contexts of $N$ inputs \cite{katharopoulos2020transformers}. As the ambition to process longer sequences and build larger, more comprehensive models grows, this quadratic bottleneck becomes increasingly prohibitive. The situation is further exacerbated by the trend towards ever-increasing model sizes and sequence lengths in pursuit of enhanced performance and generalization capabilities. 

To address these challenges, exploring specialized hardware accelerators, such as Intel Gaudi processors \cite{medina2020habana}, AMD Versal ACAP AI Engines (AIEs), SambaNova Reconfigurable Dataflow Units (RDUs) \cite{emani2021accelerating}, and Cerebras's wafer-scale engine (WSE) \cite{lie2022cerebras}, has emerged as a promising avenue for mitigating the computation demands of training and inference of large Transformer-based models. Gaudi processors stand out for their innovative architecture designed specifically to accelerate deep learning (DL) workloads and offer a heterogeneous compute architecture comprising a Matrix Multiplication Engine (MME) and a cluster of fully programmable Tensor Processing Cores (TPCs). This combination allows Gaudi to efficiently handle various DL operations, both matrix-based and non-matrix-based, with high performance and flexibility. 

Despite the potential showcased by Gaudi, Softmax operations in Transformers become a performance bottleneck when processing long sequence inputs \cite{zhang2023benchmarking}. The main reasons are \ding{182} The computational complexity of Softmax operations in a Transformer is $\mathcal{O}(N^2)$. \ding{183} Softmax operations are mapped into TPC, but reduction operations in Softmax are not well-suited for single instruction multiple data (SIMD) architectures like TPC (see more details about TPC architecture in Section \ref{sec:background}). Long sequences further exacerbate this problem especially when the sequence length exceeds 2048. \uline{Overall, the limited computational capability of TPC combined with complexities of Softmax operations in Transformers hinders Gaudi's overall performance and efficiency.}

Existing algorithmic approaches to optimize Transformers fall into three categories: \ding{182} Exploiting the sparsity of attention matrices, exemplified by methods such as Reformer \cite{kitaev2020reformer} and Big Bird \cite{zaheer2020big}. \ding{183} Applying kernel methods \cite{hofmann2008kernel}, including Performer \cite{choromanski2020rethinking} and Transformers as RNNs \cite{katharopoulos2020transformers}, to approximate and eliminate Softmax operations. This attention mechanism is referred to as \textit{linear attention} because its complexity reduces to $\mathcal{O}(N)$ upon the removal of Softmax. \ding{184} Combining diverse attention methods to enhance Transformers' performance. An example is \cite{chen2021scatterbrain}, which successfully integrates sparse and kernel methods to improve model quality. While such integrations often lead to models of higher quality, they may also result in increased computational load and slower processing speeds.

Moreover, challenges arise when directly adapting these efficient Transformer techniques to Gaudi processors. Specifically, Gaudi processors feature a heterogeneous compute architecture comprising Matrix Multiplication Engines (MME) and Tensor Processing Cores (TPC). However, the sparse attention mechanism, which introduces irregular memory access and computation, is primarily mapped onto TPCs, leaving MMEs, which are not programmable and only support dense matrix-matrix operations, idle in scenarios requiring sparse attention. Conversely, linear attention, which is fundamentally based on matrix multiplication, can utilize almost all calculations on MMEs due to their stronger computational capabilities, but this leaves TPCs idle in such cases. This situation raises a critical question: \textit{Can we effectively combine sparse and dense attention mechanisms in a way that fully leverages both MME and TPC, while maintaining the quality of the model?}

To this end, we propose an optimized \underline{G}audi-based Trans\underline{former} (called GFormer) for large language models (LLMs) acceleration on the Gaudi processor. GFormer synergistically combines sparse and linear attention mechanisms to enhance computational efficiency by fully leveraging both MME and TPC while preserving model quality. Key components of our framework include:
\ding{182} The integration of diverse attention mechanisms to optimize both computation efficiency and model fidelity.
\ding{183} The implementation of a windowed local-context self-attention kernel utilizing the vector units in TPC, aimed at maximizing computational throughput.
\ding{184} The development of an efficient outer product TPC kernel for handling a subset of the outer product operations in causal linear attention, effectively balancing the workload between MME and TPC.
\ding{185} The introduction of an optimal workload partitioning algorithm to ensure balanced utilization of TPC and MME resources.
To the best of our knowledge, \textit{this is the first work that facilitates high-performance and high-utilization LLM inference on heterogeneous hardware like Gaudi processors}. This exploration aims to harness the computational capabilities of Gaudi and the characteristics of LLMs, inspiring future innovative ML hardware designs.

The main contributions of this paper are summarized below:
\begin{itemize}
\item We introduce an innovative approach to integrate disparate sparse and linear attention mechanisms. This strategy is designed to fully utilize the computational capabilities of MME and TPC on the Gaudi processor.
\item We develop a windowed local-context self-attention kernel that is specifically tailored for TPC. This kernel is optimized to leverage TPC's local memory and vectorized load and store operations.
\item We present an efficient outer product kernel for TPC, employing the vector unit (SIMD) to optimize the processing of causal linear attention operations.
\item We introduce a performance modeling technique for TPC and MME. This model is instrumental in balancing workloads between TPC and MME.
\item We evaluate GFormer on GPT and ViT models and find that it achieves up to 2$\times$ and 2.2$\times$ speedups, respectively.
\end{itemize}

\section{Background and Motivation}
\label{sec:background}
In this section, we present background information for Transformers, the Gaudi processor architecture, the TPC programming model, and our motivation.

\subsection{Transformers}
\begin{figure}[t]
    \centering
    \includegraphics[width=0.7\linewidth]{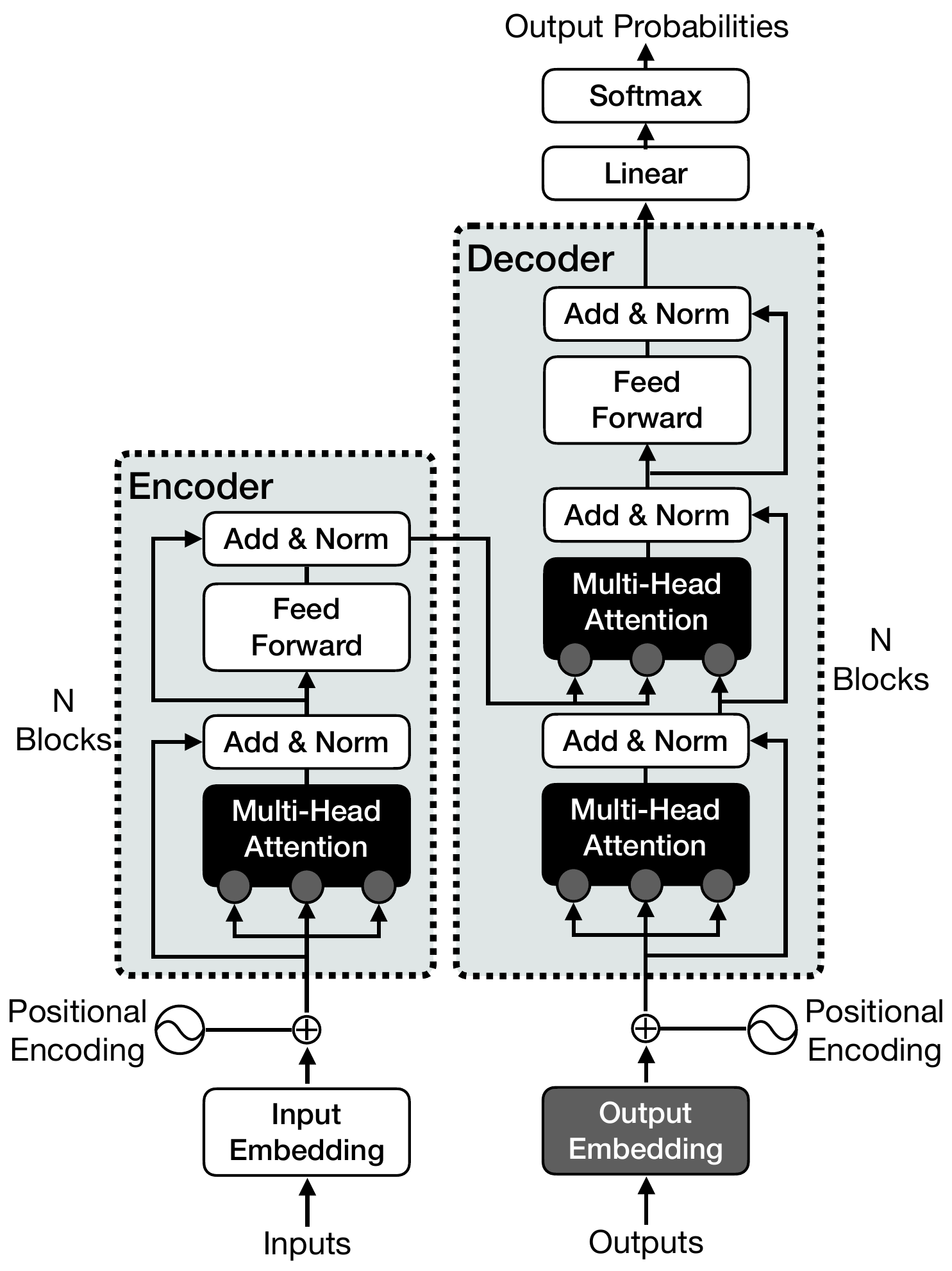}
    \vspace{-2mm}
    \caption{Overview of Transformer architecture.}
    \label{fig:transformer}
   \vspace{-2mm}
\end{figure}

Transformer \cite{vaswani2017attention} architecture departs from previous sequence-to-sequence models by relying on self-attention to draw global dependencies among inputs, which allows it to handle sequences of data in parallel and capture long-range dependencies more effectively. Figure \ref{fig:transformer} presents the architecture of a Transformer, which typically consists of an encoder, a decoder, and other operations such as position embedding. We describe the decoder for simplicity. A decoder contains masked multi-head self-attention and a fully connected feed-forward network. The masked multi-head self-attention mechanism prevents the current token from attending to tokens in masked positions. The feed-forward network provides further transformation of the attention-aggregated information.

\textbf{Causal language models} are just concerned with the previous context (tokens on the left) when predicting the next token in a sequence of tokens. In Softmax-based self attention, we add an attention mask matrix into the raw attention matrix to mask attention on the right of the current position. All elements of the lower triangular part of the attention mask matrix are 0, and the other elements of the attention mask matrix are set to $-\infty$. Here we refer to such an attention mechanism as \textbf{causal attention}. Generative Pre-trained Transformer (GPT) models \cite{radford2018improving} are based on the Transformer decoder architecture. GPT models are typical causal language models. GPT models are characterized by their large scale, extensive pre-training on diverse text corpora, and their ability to adapt to a wide range of tasks with minimal task-specific modifications. 

\textbf{Vision Transformer} (ViT) \cite{dosovitskiy2020image} adapts the Transformer for image classification tasks. By treating images as sequences of patches (akin to words in a sentence), ViT applies the \textbf{self-attention} mechanism across these patches to capture global dependencies within the image. ViT can attend to image tokens bidirectionally (full access to the image tokens on the left and right). This approach has demonstrated competitive or superior performance to conventional convolutional neural networks (CNNs) on image classification benchmarks.

\subsection{Efficient Attention Mechanism}
Efficient attention mechanisms aim to reduce the computational complexity traditionally associated with the Softmax-based attention in Transformers, which scales quadratically with the sequence length. Specifically, sparse attention selectively focuses on a subset of key positions for each query in the sequence, rather than attending to all positions. This selective focus drastically reduces the number of computations and memory requirements, as the attention matrix is no longer fully dense but sparse. For example, Longformer \cite{beltagy2020longformer} introduces a sparse attention mechanism that judiciously selects a subset of positions to attend to, blending local with a few global attention patterns. Big Bird \cite{zaheer2020big} incorporates a unique mix of random, global, and sliding window attention. This approach not only maintains the model's ability to grasp complex dependencies over vast stretches of text but also does so with enhanced flexibility and efficiency.

The equation $\text{softmax}(QK^T) V \approx \phi(Q) (\phi(K)^T V)$
expresses the idea of linear attention. Linear attention first uses the kernel method $\phi$ to project query $Q$ and key $K$ matrices into feature spaces to approximate and remove Softmax operations. We can first compute $(\phi(K)^T V)$ to avoid explicit calculation of attention matrix $\text{softmax}(QK^T)$. Thus the computation complexity of linear attention becomes $O(N)$. For example, ``Transformers are RNNs'' \cite{katharopoulos2020transformers} employs a simple feature map defined below: 
\begin{equation} 
\small
\phi(x) = elu(x) + 1
\end{equation}
The Performer \cite{choromanski2020rethinking} employs a randomized feature map to approximate the Softmax attention. 

\subsection{Gaudi Processor Architecture}
\label{subsec:gaudi}
\begin{figure}[ht]
    \centering
    \includegraphics[width=0.7\linewidth]{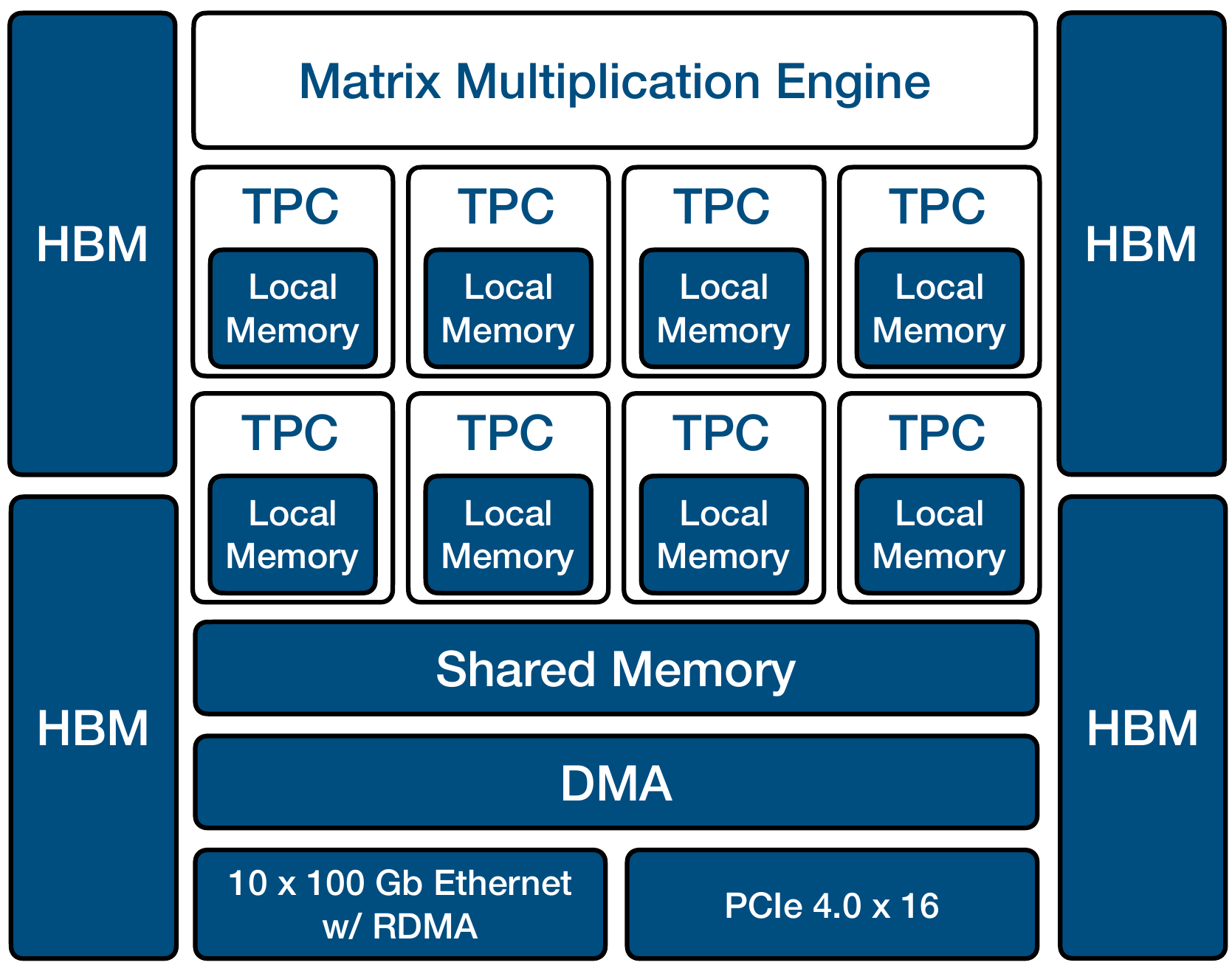}
    \vspace{-2mm}
    \caption{A high-level view of Gaudi architecture, which consists of Matrix Multiplication Engine (MME), Tensor Processing Cores (TPC), Memory Units (Local Memory, Shared Memory, DMA, HBM), and Connection Units (Ethernet, PCIe).}
    \label{fig:overview}
   \vspace{-2mm}
\end{figure}

Gaudi processor is a specialized hardware accelerator designed for deep learning training workloads \cite{medina2020habana}. As shown in Figure \ref{fig:overview}, it features a heterogeneous compute architecture with a Matrix Multiplication Engine (MME), eight fully programmable Tensor Processing Cores (TPC), and fast memory and network units \cite{HabanaWhitePap}. The MME is specifically tuned for doing all operations that can be lowered to matrix multiplication, such as fully connected layers, convolutions, and batched GEMM. The TPC is a very long instruction word (VLIW) single instruction multiple data (SIMD) processor crafted for deep learning nonlinear operations. 

The fast memory and network units enhance intra-/inter- processor data transfers. Four high-bandwidth memory (HBM) devices provide 32 GB of capacity with one terabyte-per-second of memory bandwidth. Shared memory can be used to streamline the data exchange between MME and TPC. On-chip ten 100 gigabit integrated remote direct memory access (RDMA) over converged Ethernet (RoCE) ports facilitate efficient inter-processor communication.

\subsection{TPC Programming}
\paragraph{\textbf{TPC architecture}}
TPC is responsible for executing non-linear deep learning operators. Its wide SIMD vector unit supports 2048-bit SIMD operations with data types such as float, bfloat16, INT16, INT32, and INT8. The TPC’s arithmetic logic unit can execute up to 64 floats/INT32, 128 INT16, or 256 INT8 operations per cycle. Multiple TPC cores in Gaudi can be executed in parallel. 

TPC processor includes four distinct memory spaces: scalar local memory, vector local memory, global memory, and configuration space. Global memory is accessed through specialized access points termed tensors. A 2,048-bit vector can be loaded from or written to global memory every four cycles, on average. Local memory of each TPC processor is divided into scalar local memory (1 KB) and vector local memory (80 KB). Local memory can be either read from or written to on every cycle with no bandwidth constraint \cite{TPCProgramming}.

\paragraph{\textbf{TPC programming}}
TPC is programmed via TPC-C, a derivative of C language. A TPC program contains TPC code (kernel) and host glue code. TPC code is the actual kernel implementation. TPC CLANG compiler is based on LLVM and is used for TPC kernels' compilation, simulation, and debugging. Host glue code is executed on the host machine and controls TPC kernels' execution. A TPC kernel only accepts tensors as inputs or outputs with dimensions ranging from 1 to 5. Index spacing, similar to threads in CUDA programming, efficiently divides workloads among TPC processors. Each index space member corresponds to an independent unit of work executed on a single TPC. TPC CLANG compiler also provides intrinsic functions for optimized kernel implementation. Intrinsics encompass arithmetic, bitwise, logical, load, store, et al, operations. 

\begin{figure}[t]
    \centering
    \includegraphics[width=0.8\linewidth]{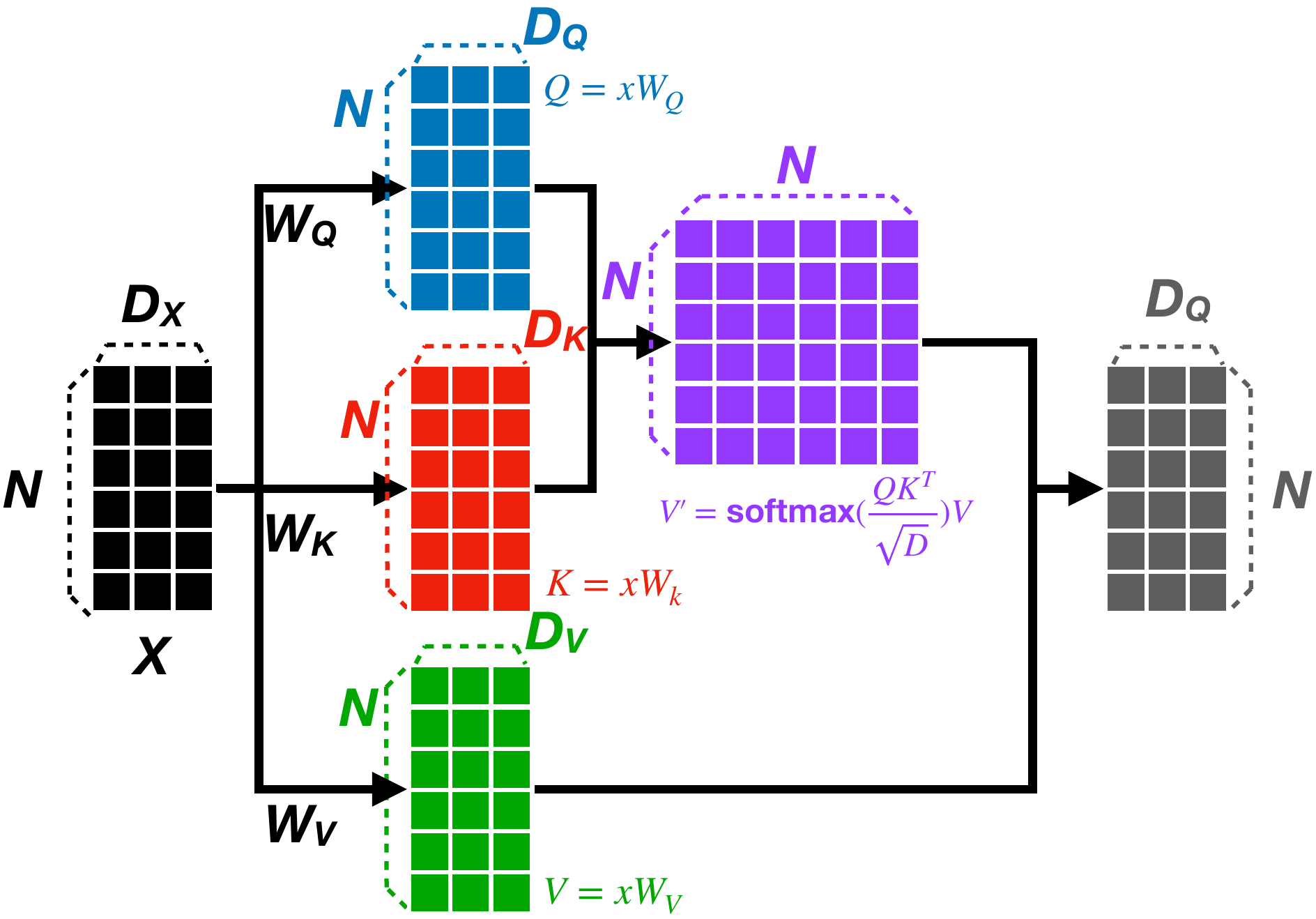}
    \vspace{-2mm}
    \caption{Matrix Computation workflow of each self-attention. $Q$, $K$ and $V$ are query, key, value matrices of dimension size $N$ by $D_Q$,$D_K$, $D_V$, respectively.}
    \label{fig:matrix_computation}
   \vspace{-2mm}
\end{figure}

\subsection{Motivation}
The impressive ability of Transformer-based models comes from complex computational operations and the huge number of parameters (340 million in BERT, 1.5 billion in GPT-3) \cite{devlin2018bert,brown2020language}, which results in intensive computations during training and inference. Consequently, training and inference of Transformer-based models is both time-consuming and resource-intensive. Utilizing a new efficient Transformer architecture is a possible solution to reduce computation complexity. However, the Gaudi-specific optimizations on Transformer architecture are not well studied. Additionally, Figure \ref{fig:matrix_computation} shows the computation flow of Softmax-based self-attention. Specifically, The input sequence $x \in \R ^{N \times D_x}$ is projected by three weight matrices $W_Q, W_K, W_V$ to corresponding representations $Q$, $K$ and $V$. Following common terminology, the $Q$, $K$, and $V$ are referred to as the "queries", "keys", and "values" respectively. Then Softmax is used to normalize the attention matrix $QK^T$ into a probability distribution. As indicated in \cite{zhang2023benchmarking}, The Softmax operation is only executed on TPC and becomes a performance bottleneck when processing long sequence inputs. But there is no existing approach to breaking this bottleneck on Gaudi processors. Furthermore, Gaudi processors feature heterogeneous compute architecture comprising MME and TPC. It is worthwhile to balance workloads between MME and TPC to fully utilize the computation resources of both MME and TPC. However, there is no previous method to investigate balancing workloads on Gaudi processors. 

{\small
\begin{align}
    \begin{aligned}
    Q & = x W_Q \\
    K & = x W_k \\
    V & = x W_V \\
    V^{\prime} & = \text{softmax}(\frac{QK^T}{\sqrt{D}})V
    \end{aligned}
    \label{equ-1}
\end{align}
}

\begin{figure*}[t]
    \centering
    \includegraphics[width=1.0\linewidth]{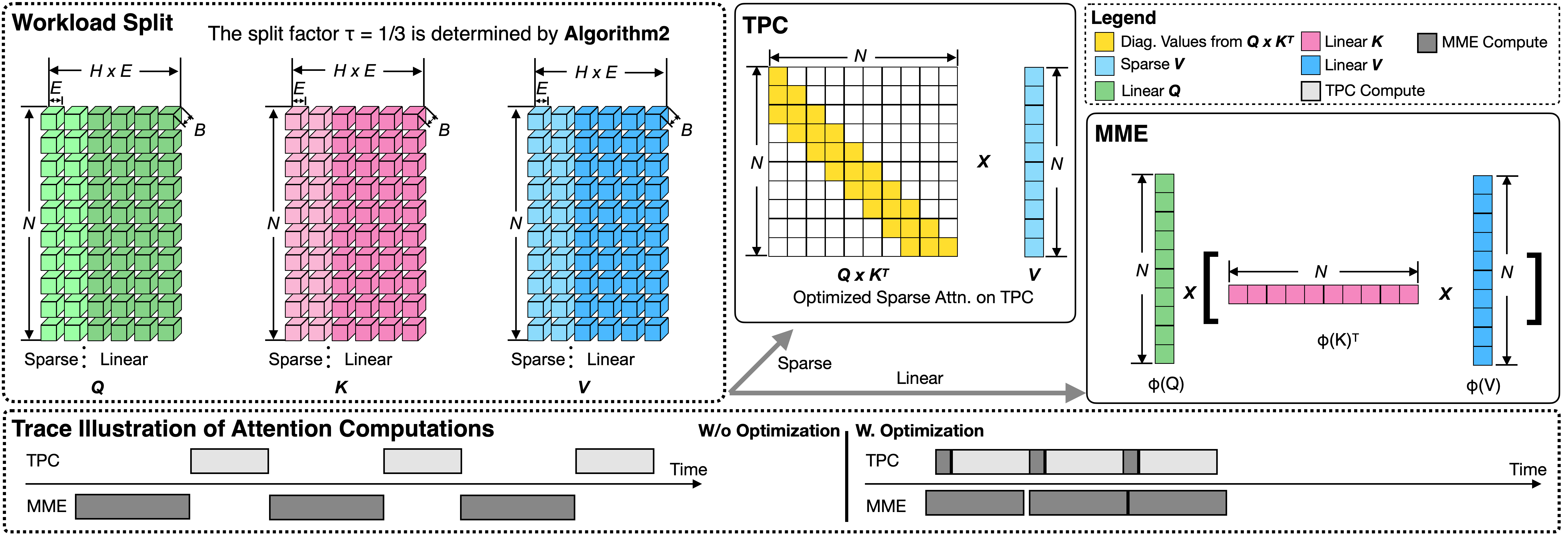}
    \vspace{-2mm}
    \caption{Overview of GFormer workflow. $\tau$ of the heads is for sparse attention.
    }
    \label{fig:Overview}
    \vspace{-4mm}
\end{figure*}

\section{Design Methodology}
\label{sec:design}
In this section, we propose our optimized Transformer design and optimized TPC kernels. 

\subsection{Overview of This Work}
\label{subsec:overview}
Figure~\ref{fig:overview} shows the overview of GFormer. The Gaudi processor is a heterogeneous architecture comprising a cluster of TPCs, as well as configurable MMEs. To maximize the utilization of both TPC and MME, our proposed design effectively combines sparse and linear attention approximations in the following ways. Inputs of self-attention are $Q$, $K$, and $V$. They are referred to as the “queries”, “keys” and “values” respectively. $Q, K, V \in \R ^{B \times N \times H \times E}$, where $B$, $N$, $H$, and $E$ are batch size, sequence length, the number of heads, head size, respectively. We split $Q, K, V$ along the head dimension ($H$) into two groups, as inputs of sparse attention and inputs of linear attention. The partition is according to a hyperparameter $\tau$. $H \times \tau$ is the number of heads for sparse attention and $H \times (1-\tau)$ is the number of heads for linear attention. 

For the sparse attention part, to take full advantage of capabilities of SIMD architecture in TPC, we adopt a windowed local-context self-attention and implement an efficient TPC kernel. Window attention avoids irregular data access and enables task partition across multiple TPCs. For the linear attention part, inspired by the Performer, we use positive orthogonal random features to approximate the Softmax operation in the Transformer \cite{choromanski2020rethinking}. Specifically, $\text{softmax}(QK^T) V \approx \phi(Q) (\phi(K)^T V) $, where $\phi$ is the feature map. Most calculations of linear attention are matrix-matrix multiplication and can be mapped to MME, which brings two benefits. (1) it takes advantage of powerful MME. (2) it avoids the data movement between MME and TPC. Our mixed approach not only maximizes hardware utilization in the Gaudi processor but also helps reduce the accuracy loss caused by the Softmax approximation. We expect that executions of TPC and MME will overlap through our optimization.

\subsection{TPC Best Fitted Sparse Attention}
\label{subsec:sparse}
\textit{\textbf{Problems}}:
Softmax applies the standard exponential function to each element of the input tensor and normalizes these values by dividing by the summation of all these exponentials along a specific dimension. The computational complexity of Softmax operations is $\mathcal{O}(N^2)$. Sparse attention, for example, Longformer \cite{beltagy2020longformer}, Big Bird \cite{zaheer2020big}, is proposed to reduce computational complexity. 

\textit{\textbf{Challenges}}:
However, we face two challenges when performing Softmax on an irregularly sparse attention matrix. First, irregular data access leads to high data reading latency and low TPC utilization. Second, it is not able to perform index space mapping (divide workloads) between TPC processors evenly. The TPC programming only supports linear transformations for mapping, but the random number of non-zero elements in each row of the sparse attention matrix causes these linear transformations to fail. 

\textit{\textbf{Proposed design}}:
To overcome these challenges and benefit the computation pattern of TPC, we adopt a windowed local-context self-attention. Our attention pattern employs a fixed-size window attention surrounding each token. Using multiple stacked layers of such windowed attention results in a large receptive field, where top layers have access to all input locations and have the capacity to build representations that incorporate information across the entire input, similar to CNNs. Given a fixed window size $w$, each token attends to $w$ previous (left) tokens. Besides, the window size is a multiple of 64 to fully utilize vector units in TPC. Since a TPC's wide SIMD vector unit supports 2048-bit SIMD operations. The TPC can execute up to 64 float operations in parallel in one cycle. Additionally, the TPC prefers directly loading a vector of values from global memory. 

Listing \ref{lst:alg1} illustrates the pseudocode of windowed attention on TPC. Specifically, we declare exp\_x in local memory (Line 1) to enable high-bandwidth reading and writing. We directly load a vector of data into x from the input tensor (Line 22). We apply the exponential function to each element of x to generate y (Line 23). We store y into exp\_x (Line 24). We avoid writing y into global memory to speed up intermediate data write and read since window size is typically less than 256, in which all exponential data y can be held in local memory. Then we obtain reciprocal of the sum of exponents and multiply all exponents by the reciprocal value (Lines 28-40). The advantages of this TPC kernel are (1) utilizing local memory. (2) vectorized load and store. 

\begin{lstlisting}[style=C-style, linewidth=\linewidth, caption=Pseudocode of windowed attention on TPC., captionpos=b, label=lst:alg1, numberstyle=\tiny\color{red}]
__local__ float64 exp_x[C];
void main(tensor in, tensor out, int window_start, int window_size) {
const int5 index_space_start = get_index_space_offset();
const int5 index_space_end = get_index_space_size() + index_space_start;
const int width_step  = 64;
int w_s = window_start;
int w_e   = window_start + window_size;
const int height_step  = 1;
const int h_s=index_space_start[1]*height_step;
const int h_e=index_space_end[1]*height_step;
int5 coords = {w_s, h_s, 0, 0, 0};
float64 x, y, sum;
int e_i = 0;
for (int h = h_s; h < h_e; h += height_step)
{
  coords[1] = h;
  sum = 0.f;
  for (int w = w_s; w < w_e; w += width_step)
  {
    coords[0] = w;
    // Load input tensors
    x = v_f32_ld_tnsr_b(coords, in);
    y = v_exp_cephes_fast_f32(x);
    exp_x[e_i] = y;
    e_i = e_i + 1;
    sum = sum + y;
  }
 // Sum across the vector
  sum = v_f32_reduce_add(sum);
 // 1/(sum_of_exponents)
  sum = v_reciprocal_f32(sum);
  e_i = 0;
  for (int w = w_s; w < w_e; w += width_step)
  {
    coords[0] = w;
    x = exp_x[e_i];
    e_i = e_i + 1;
    // Multiply exp(x) * 1/(sum_of_exponents)
    y = x * sum;
    v_f32_st_tnsr(coords, out, y);
  }
} }
\end{lstlisting}

\subsection{Efficient Outer Product on TPC}
\label{subsec:outer}
\textit{\textbf{Problems}}:
For causal linear attention, we need manually to let the current token only pay attention to previous tokens since there is no attention mask matrix in the causal linear attention scenario. Algorithm \ref{alg:causal} describes computation procedures in causal linear attention. Inputs of the procedure are $Q$, $K$, and $V$. The $Q$, $K$, and $V$ are referred to as the “queries”, “keys” and “values” respectively. The output of the procedure is $H$ (hidden state). Given that subscripting a matrix with $i$ returns the $i$-th row as a vector. Within the loop, each pair of $Q_i$ and $K_i$ is transformed using a random feature map $\phi$, resulting in ${Q}_i^\prime$ and ${K}_i^\prime$. A normalized output vector $H_i$ is computed via accumulation of outer product of ${K}_i^\prime$ and $V_i$ from $1$ to $i$. 

\begin{algorithm}[h]
\scriptsize\sffamily
\normalfont
\SetAlgoLined
\setcounter{AlgoLine}{0}
\SetKwComment{Comment}{\# }{}
\SetKwInOut{Input}{Inputs}
\SetKwInOut{Output}{Outputs}
\Input{$\{Q_i\}_{i=1}^{N}$, $\{K_i\}_{i=1}^{N}$, $\{V_i\}_{i=1}^{N}$}
\Output{$\{H_i\}_{i=1}^{N}$}
\BlankLine
$A \in \R ^{E \times E}$, $Z \in \R ^E$ \par
$A, Z \gets 0, 0$ \par
\For{$i=1$ \KwTo $N$}{
    \textcolor{Green}{\# Random feature maps} \par
    ${Q_i}^\prime, {K_i}^\prime \gets \phi(Q_i), \phi(K_i)$ \par
    $A \gets A + {K_i}^\prime \bigotimes V_i$ \par
    $Z \gets Z + {K_i}^\prime$ \par
    $H_i^T \gets {{K_i}^\prime}^{T} S / ({Q_i}^\prime \cdot Z)$ \par
}
\caption{\footnotesize Causal linear attention.}
\label{alg:causal}
\end{algorithm}

\begin{figure}[h]
    \centering
    \includegraphics[width=1.0\linewidth]{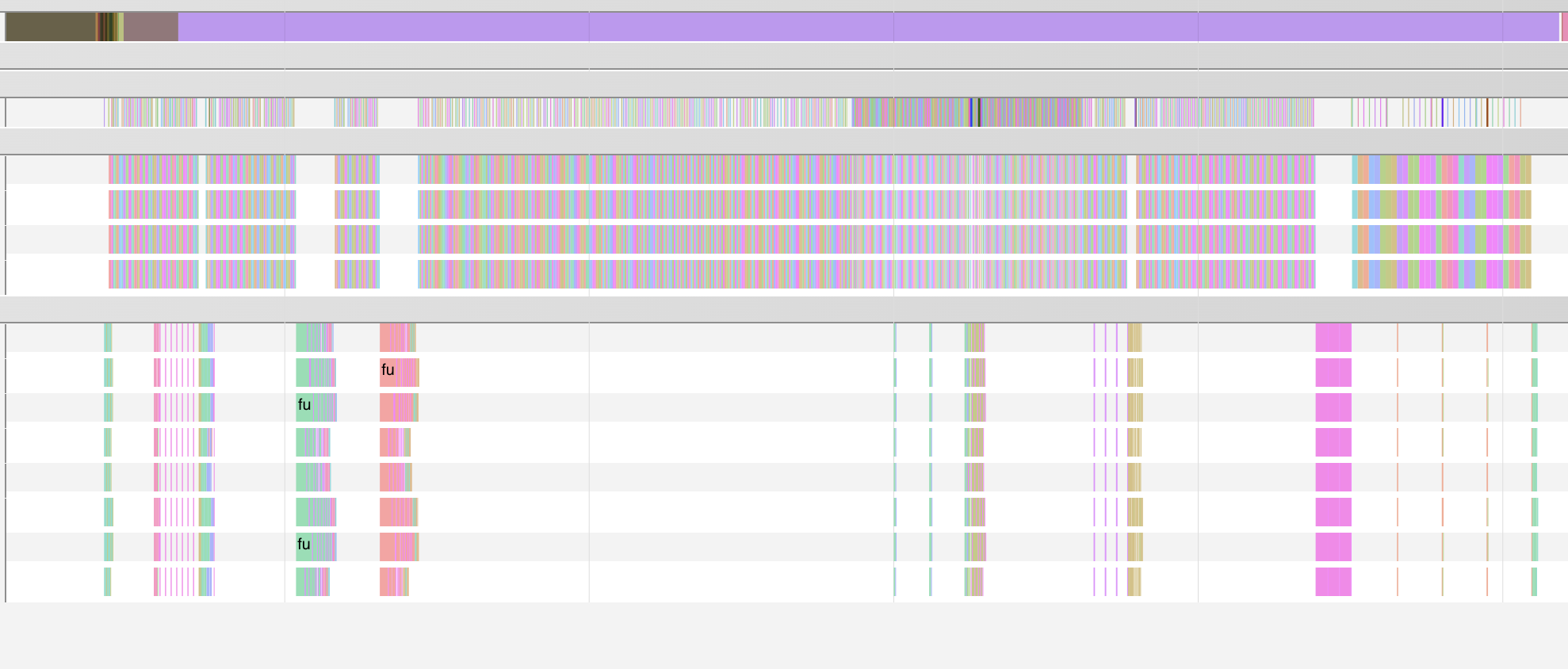}
    \vspace{-2mm}
    \caption{Profiling result of original causal linear attention.}
    \label{fig:ori_out}
   \vspace{-4mm}
\end{figure}

Listing \ref{lst:alg2} describes the Pytorch implementation of the causal linear attention on Gaudi. Specifically, MME is not programmable and only supports matrix-matrix multiplication. To achieve outer product operation, we first insert a dimension of size one into $k$ and $v$ at the specified dimension. Then we perform a batch matrix-matrix multiplication of $k\_prime$ and $v$. We then perform the cumulative sum of $att\_raw$ and $att\_norm$ over the sequence length dimension. Figure \ref{fig:ori_out} shows the profiling result of causal linear attention of such implementation. We can find the MME is overwhelmed by matrix-matrix multiplication. But TPC is quite idle. The reason is that the last dimensions of $k$ and $v$ (head dimension) are usually less than 64, Lines 7 and 8 in Listing \ref{lst:alg2} are actually $B \times H \times N$ small matrix-matrix multiplication ($1 \times E \times E \times 1$), where $B$, $H$, $N$, and $E$ are batch size, the number heads, sequence length, head size, respectively. The small size of matrix-matrix multiplication is not very efficient in MME.  

\begin{lstlisting}[style=python-style, linewidth=\linewidth, caption=Pseudocode of causal linear attention., captionpos=b, label=lst:alg2, numberstyle=\tiny\color{red}]
def linear_causal_attention(q, k, v):
    # Project key and queries onto the feature map space
    k_prime = feature_map(k)
    q_prime = feature_map(q)

    ref_v = torch.ones_like(v.unsqueeze(2))
    out_k_v=k_prime.unsqueeze(3)@v.unsqueeze(2)
    norm = k_prime.unsqueeze(3) @ ref_v
    # Consolidate against the feature dimension
    att_raw = q_prime.unsqueeze(2) @ out_k_v
    att_norm = q_prime.unsqueeze(2) @ norm
    # Cumulative sum over the sequence
    att_raw = att_raw.cumsum(1)
    att_norm = att_norm.cumsum(1)
    att_raw = att_raw.squeeze(2)
    att_norm = att_norm.squeeze(2)
    # Normalize
    attn = att_raw / att_norm
    return attn
\end{lstlisting}

\textit{\textbf{Challenges}}:
To balance the utilization of both MME and TPC. We propose to map a proportion of outer product operations onto TPC. However, the challenge is to implement an efficient outer product kernel on TPC, which achieves similar performance as MME. Otherwise, the outer products on TPC will instead become a bottleneck. Specifically, we need to solve the following issues: \Circled{1} We need to evenly distribute outer product operations among TPC to avoid load imbalance problems. \Circled{2} As aforementioned, we need the For loop to implement $B \times H \times N$ outer product operations, but serial execution characteristics of the For loop will hinder instruction parallelism on TPC. \Circled{3} In the outer product between vector 1 and vector 2. One element from vector 1 is multiplied by all elements of vector 2. Additionally, we prefer to directly load a vector of data into TPC for higher data reading bandwidth. However, TPC intrinsics do not support accessing a specific element in the vector data, which prevents us from achieving the outer product.

\begin{figure}[h]
    \centering
    \includegraphics[width=1.0\linewidth]{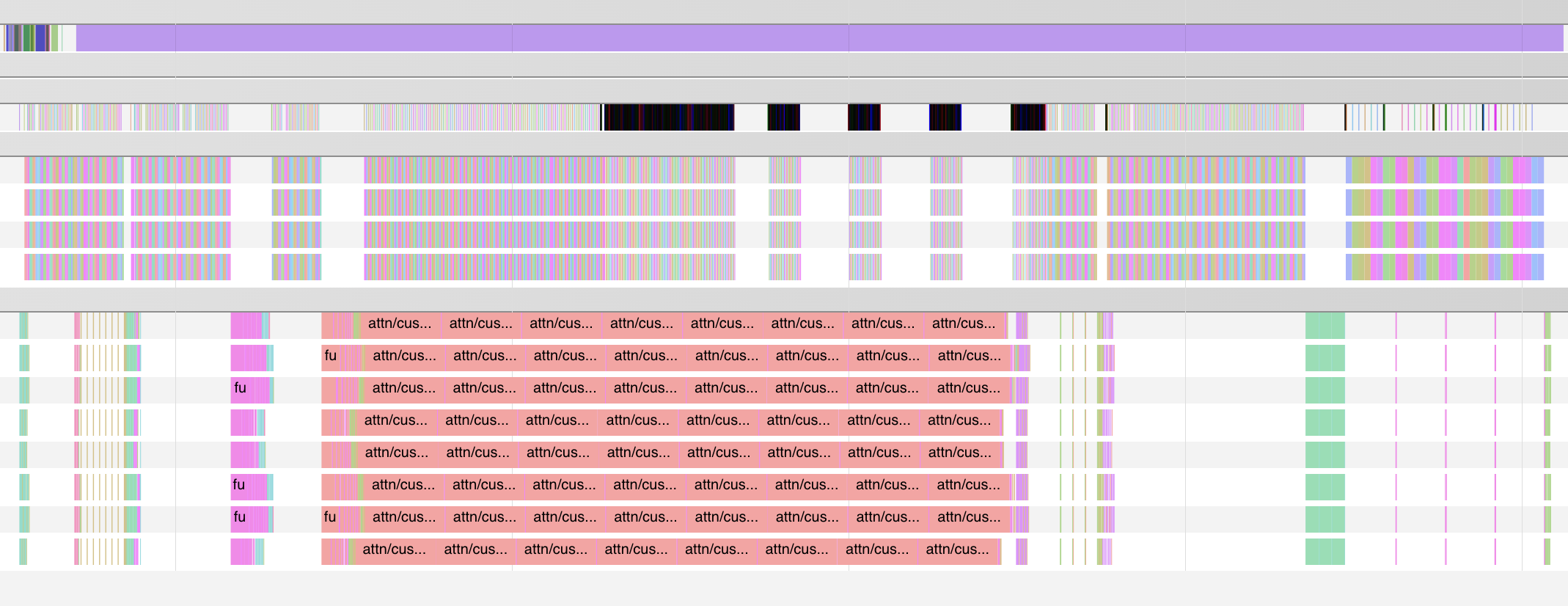}
    \vspace{-2mm}
    \caption{Profiling result of optimized causal linear attention.}
    \label{fig:opt_out}
   \vspace{-2mm}
\end{figure}

\textit{\textbf{Proposed implementation}}:
To this end, we proposed an efficient outer product kernel design in  Listing \ref{lst:alg3} to solve the aforesaid problems. Specifically, we assume the head size is 64. To solve the first issue, we partition computation using index space (Lines 3-10) to evenly distribute outer product operations among TPCs. We load a row of data from a\_mat and b\_mat into a\_v, and b\_v, respectively (Lines 17-18). For the second issue, We unroll the For loop to increase the instruction level parallelism (Line 19). For the third issue, we design a function, v\_broadcast\_element\_f32 to enable access to the element in vector and convert the element into vector. The idea of v\_broadcast\_element\_f32 is bit-level shuffle and shift. We then can broadcast the element in a\_v at w position into a vector a\_x. Then b\_v is multiplied by a\_x to generate an output vector (Line 20-24). The advantages of this TPC kernel are (1) vectorized load and store. (2) efficient vector multiplication using SIMD. (3) without insertion of an extra dimension (avoiding memory movement). Figure \ref{fig:opt_out} depicts the profiling result of causal linear attention after optimization. Both MME and TPC have relatively balanced utilization. 

\begin{lstlisting}[style=C-style, linewidth=\linewidth, caption=Pseudocode of outer product on TPC., captionpos=b, label=lst:alg3, numberstyle=\tiny\color{red}]
#define A_LEN 64
void main(tensor a_mat, tensor b_mat, tensor o_mat) {
  const int5 index_space_start = get_index_space_offset();
  const int5 index_space_end = get_index_space_size() + index_space_start;
  int5 a_coords = {0};
  int5 b_coords = {0};
  int5 o_coords = {0};
  const int height_step  = 1;
  const int h_s=index_space_start[1]*height_step;
  const int h_e=index_space_end[1]*height_step;
  for (int h = h_s; h < h_e; h += height_step) {
    a_coords[1] = h;
    a_coords[0] = 0;
    b_coords[1] = h;
    o_coords[1] = h;
    // Load a row from a_mat, b_mat
    float64 a_v=v_f32_ld_tnsr_b(a_coords, a_mat);
    float64 b_v=v_f32_ld_tnsr_b(b_coords, b_mat);
    #pragma unroll (8)
    for(int w = 0; w < A_LEN; w += 1) {
     float64 a_x=v_broadcast_element_f32(a_v, w);
      float64 out_v = v_f32_mul_b(a_x, b_v);
      o_coords[0] = w * A_LEN;
      v_f32_st_tnsr(o_coords, o_mat, out_v);
    }
  } }
\end{lstlisting}

\subsection{Optimal Partition Algorithm}
\label{subsec:partition}
As discussed in Section \ref{subsec:overview}, to fully utilize both MME and TPC, we partition $Q$, $K$, and $V$ along head dimension according to a hyperparameter $\tau$ for sparse attention and linear attention. The overall system's runtime is determined by the maximum runtime between the TPC and MME, so it is crucial to balance their respective workloads. We are required to carefully determine the hyperparameter $\tau$ to make MME and TPC has balanced workloads. To achieve this balance, we model computation latencies of both the TPC and MME, which allows us to obtain the relationship between workload and runtime for each engine. By estimating the runtime of TPC and MME based on their respective workloads, we can obtain the optimal partition when their runtime is equal or similar. 

Specifically, we first analyze workloads' floating point of operations (FLOPs) quantitatively. We then develop a set of micro-benchmarks to measure the average performance of workloads on the TPC and MME. These micro-benchmarks cover a wide range of sizes and sparsity levels. Algorithm \ref{alg:partition} reveals how to obtain the optimal partition using FLOPs and average performance. Inputs are initial partition (p), problem size (size), FLOPs per unit of partition 0 (FLOPs\_per0), FLOPs per unit of partition 1 (FLOPs\_per1), average performance of partition 0 (perf0), and average performance of partition 1 (perf1). The output ($p^\prime$) is an optimal partition. We estimate the computation latency of partitions 0 and 1 (Lines 3-4). We find the optimal partition when latency0 and latency1 are equal or similar (Line 5).

\begin{algorithm}[h]
\scriptsize\sffamily
\normalfont
\SetAlgoLined
\setcounter{AlgoLine}{0}
\SetKwComment{Comment}{\# }{}
\SetKwInOut{Input}{Inputs}
\SetKwInOut{Output}{Outputs}
\Input{p, size, FLOPs\_per0, FLOPs\_per1, perf0, perf1}
\Output{$p^\prime$}
\BlankLine
FLOPs\_p0 = p $\times$ size $\times$ FLOPs\_per0 \par
FLOPs\_p1 = (1-p) $\times$ size $\times$ FLOPs\_per1 \par
latency0 = $\frac{\textrm{FLOPs\_p0}}{\textrm{perf0}}$ \par
latency1 = $\frac{\textrm{FLOPs\_p1}}{\textrm{perf1}}$ \par
\textrm{Find} $p^\prime$ \textrm{let} latency0 = latency1
\caption{\footnotesize Optimal partition algorithm.}
\label{alg:partition}
\end{algorithm}
\vspace{-2mm}

By estimating and balancing their runtime, we can achieve better overall performance on the Gaudi processor. Specifically, assuming $H_0$ is the number of sparse heads, $H_1$ is the number of linear heads, and $\frac{H_0}{H_1} = \frac{\tau}{(1-\tau)}$. From the perspective of linear attention, the number of FLOPs for applying random features onto $Q$ and $K$ is $4BN(H_1 E)^2$. The number of FLOPs for $C = \phi(K)^T V$ is $2BN(H_1 E)^2$. The number of FLOPs for $ \phi(Q) C$ is $2BN(H_1 E)^2$. Then the number of FLOPs for linear attention is: 
\begin{equation} 
\small
4BN(H_1 E)^2 + 2BN(H_1 E)^2 + 2BN(H_1 E)^2 = 8BN(H_1 E)^2
\end{equation}
Thus the computation latency for linear attention on MME can
be estimated as
\begin{equation} 
\small
MME\_latency = \frac{8BN(H_1 E)^2}{MME\_perf}
\end{equation}

From the perspective of sparse attention, the number of FLOPs for 
$R = QK^T$ is $2BH_0 E N W$, where $W$ is the window size. The number of FLOPs for sparse Softmax $A = \text{sparse softmax}(R)$ operations is $3BH_0 N W$, where $W$ is the window size. 
Then the number of FLOPs for sparse attention is: 
\begin{equation} 
\small
2BH_0 E N W + 3BH_0 N W = 5BH_0 N W
\end{equation}
Thus the computation latency for sparse attention on TPC can
be estimated as
\begin{equation} 
\small
TPC\_latency = \frac{5BH_0 N W}{TPC\_perf}
\end{equation}
We can obtain a suitable partition $\tau$ when letting $TPC\_latency$ is close to $MME\_latency$.

\section{Performance Evaluation}
\label{sec:evaluation}
In this section, we present our experimental setup and demonstrate the effectiveness of GFormer compared with other solutions using different models.

\subsection{Experimental Setup}
\label{subsec:setup}
\paragraph{\textbf{Platforms}}
We perform our experiments on one Habana Labs System 1 (HLS-1) \cite{medina2020habana} AI training system. We implement HLS-1 using AWS EC2 DL1 instances \cite{kokkinos2013cost}. The HLS-1 incorporates eight Gaudi processors and two Gen 4.0 PCIe switches. External host CPU is used to manage HLS-1 via PCIe switches. Each Gaudi processor is equipped with 1 MME, 8 TPC, and 32 GB on-chip memory. All experiments are on a single Gaudi processor.

\paragraph{\textbf{Implementation details}}
We implement our proposed models based on PyTorch. Gaudi software stack version is 1.14.0. The Gaudi software stack includes the graph compiler, Gaudi driver, Gaudi firmware, and corresponding PyTorch. The PyTorch version is 2.1.1.

\paragraph{\textbf{Models}}
For GPT model, we adopt GPT-Neo \cite{gpt-neo} architecture in Huggingface \cite{wolf2019huggingface}. Other representative LLMs such as Llama \cite{touvron2023llama} have similar structures. So for simplicity, we just evaluate the GPT model. For ViT model, we use the Vision Transformer \cite{wu2020visual} architecture in Huggingface. We set batch size, the number of layers, the number of heads, and head size as 4, 12, 16, and 64, respectively. These are realistic scenarios of parameter configurations found in GPT-3 with 125 million parameters \cite{gpt-neo} and Vision Transformer (ViT) base model \cite{wu2020visual}.

\subsection{Evaluation on Speedup of Sparse Attention Kernel}
\begin{figure}[h]
    \centering
    \includegraphics[width=1.0\linewidth]{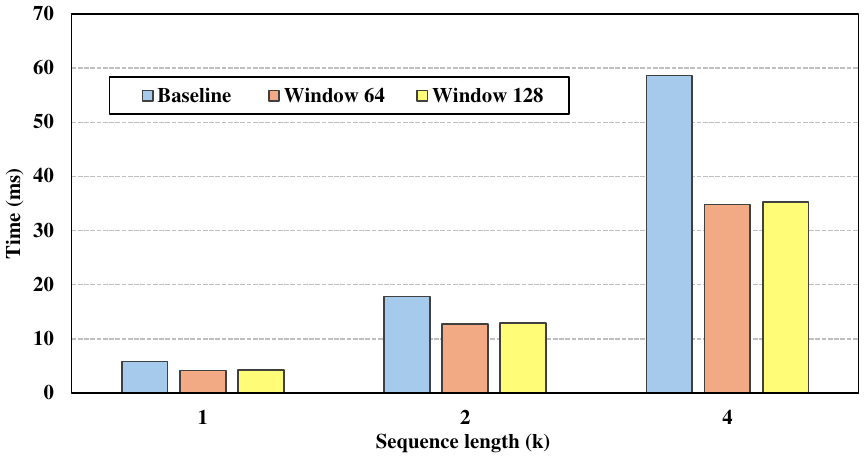}
    \vspace{-2mm}
    \caption{Speedup of windowed attention kernel.}
    \label{fig:window_attn}
   \vspace{-4mm}
\end{figure}

Figure \ref{fig:window_attn} depicts the speedup of the windowed attention kernel on different window sizes and sequence lengths. We vary sequence length from 1k to 4k. Further increasing sequence length causes memory errors. The baseline is Softmax-based attention. Window 64 and Window 128 are short for windowed attention with window sizes 64 and 128. As the sequence length increases, the speedups of windowed attention over baseline are up to 1.7$\times$.

\subsection{Evaluation on Performance of Outer Product kernel}
\begin{figure}[h]
    \centering
    \includegraphics[width=1.0\linewidth]{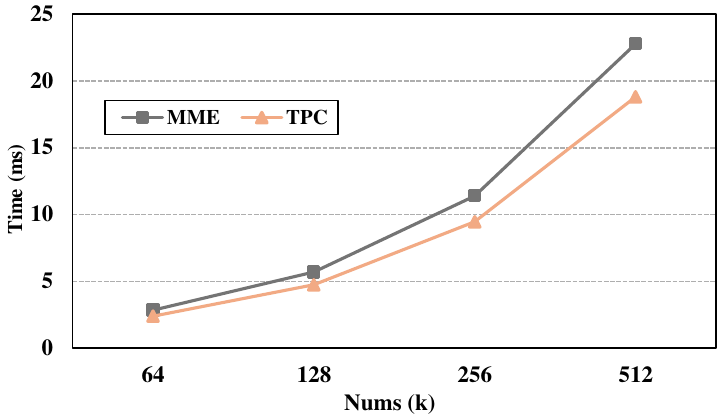}
    \vspace{-4mm}
    \caption{Performance of Outer Product Kernel.}
    \label{fig:outer_perf}
   \vspace{-2mm}
\end{figure}

As illustrated in Figure \ref{fig:outer_perf}, we compare the performance of outer product kernels on both MME and TPC. As discussed in \S\ref{subsec:outer} we use batch matrix-matrix multiplication to implement the outer product on MME. We directly implement our outer product kernel on TPC using TPC intrinsic. To compare the performance of outer product operations on CPU and GPU, we vary sequence length from 1k to 4k and the corresponding number of outer products varies from 64k to 512k. Even though TPC is less computationally powerful than the MME, the outer product kernel on TPC achieves a better performance (1.2$\times$) than the outer product on MME.

\subsection{Evaluation on Partition}
\begin{figure}[h]
    \centering
    \includegraphics[width=1.0\linewidth]{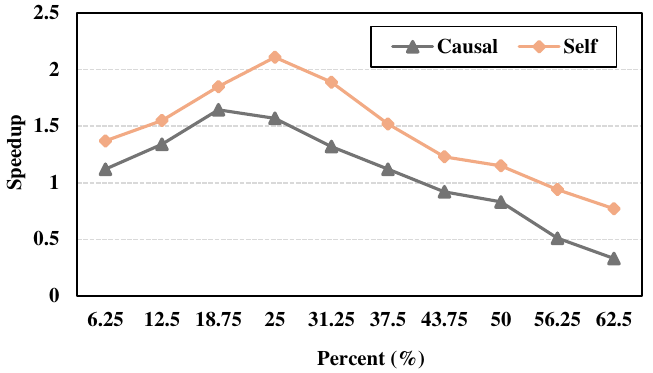}
    \vspace{-4mm}
    \caption{Performance of different partition. Causal and self are short for causal attention and self-attention.}
    \label{fig:partition}
   \vspace{-2mm}
\end{figure}

We evaluate our partition algorithm on both causal and self-attention. Here we fix the sequence length to 4k. Here we set MME\_perf and TPC\_perf as 13.37 TFLOPS and 2.31 TFLOPS according to extensive micro-benchmarks. Figure \ref{fig:partition} shows the speedup over baseline using different percentages of sparse attention. For causal attention, we find the ideal number of sparse attention heads is 3 according to our partition algorithm. The experiment also shows that there is maximum speedup when the number of sparse attention heads is 3. For self-attention, we find the ideal number of sparse attention heads is 4 according to our partition algorithm. As shown in the Figure \ref{fig:partition}. There is maximum speedup when the number of sparse attention heads is 4. This experiment proves the effectiveness of our partition algorithm. 

\subsection{Evaluation on Mixed Attention}
\begin{figure}[t]
    \centering
    \includegraphics[width=1.0\linewidth]{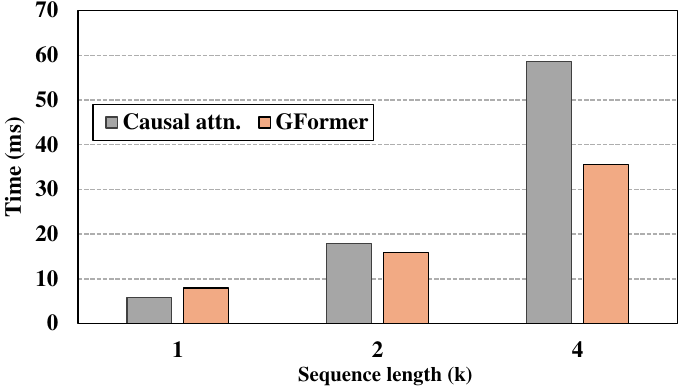}
    \vspace{-4mm}
    \caption{Performance of causal attention.}
    \label{fig:causal_attn}
   \vspace{-2mm}
\end{figure}

\begin{figure}[t]
    \centering
    \includegraphics[width=1.0\linewidth]{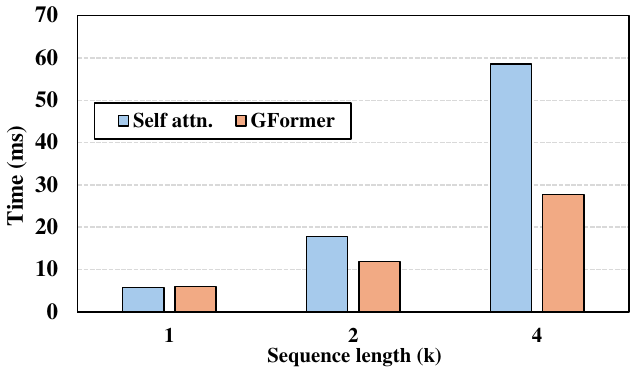}
    \vspace{-4mm}
    \caption{Performance of self attention.}
    \label{fig:self_attn}
   \vspace{-2mm}
\end{figure}

GPT or ViT is composed of multiple identical attention layers. Hence we evaluate the performance of a single attention layer across different sequence lengths as shown in Figure \ref{fig:causal_attn} \ref{fig:self_attn}. For causal attention, GFormer achieves up to 1.6$\times$ speedup over the baseline. For self-attention, GFormer achieves up to 2.1$\times$ speedup over the baseline.

\subsection{Evaluation on Speedup and Accuracy}
We evaluate our method’s speedup and approximation accuracy in LLM and ViT. We compare it with baseline, Performer, and Big Bird. The baseline uses Softmax attention. The Performer adopts linear attention. The Big Bird uses sparse attention. Our approach mixes sparse attention and linear attention. All models contain 12 layers and 16 attention heads in each layer.

\begin{figure}[h]
    \centering
    \includegraphics[width=1.0\linewidth]{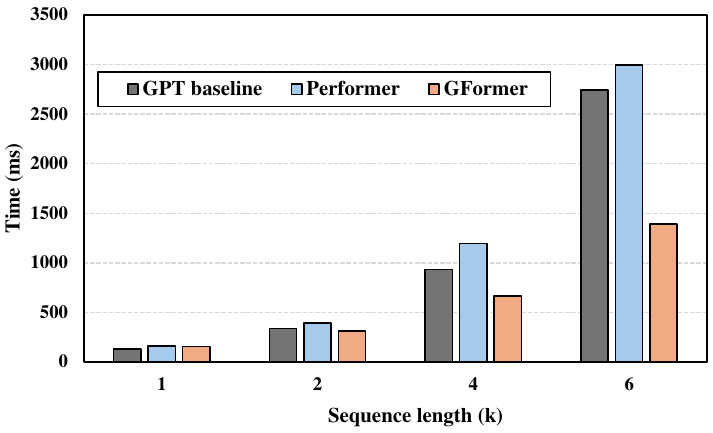}
    \vspace{-4mm}
    \caption{Performance of GPT with our attention on Gaudi.}
    \label{fig:gpt-gaudi}
   \vspace{-2mm}
\end{figure}

\paragraph{\textbf{LLM Speedup}}
Figure \ref{fig:gpt-gaudi} shows the performance of GPT with different attention mechanisms on Gaudi. Compared with the baseline, The Performer takes 16\% more run time since causal linear attention in the GPT model has low computation efficiency. GPT with GFormer achieves 2.0 $\times$ speedup over baseline when sequence length is 6k due to efficiently balancing computation of linear causal attention on both MME and TPC. 

\paragraph{\textbf{LLM Accuracy}}
As shown in Table \ref{tab:perplexity}, we compare perplexity between different models across wikitext-103 \cite{merity2016pointer} and bookcorpus \cite{Zhu_2015_ICCV}. We pre-train all models on wikitext-103 and bookcorpus for 120k steps. Lower Perplexity means better model performance. Perplexity (PPL) is one of the most common metrics for evaluating autoregressive or causal language models. Perplexity is defined as the exponentiated average negative log-likelihood of a sequence. For a tokenized sequence $X = (x_0, x_1, , x_t)$, $PPL(x) = exp(-\frac{1}{t}\sum_i^t log p(x_i|x_{<i}))$. The perplexity of our method is only 1.2 higher than the baseline but lower than other methods.

To evaluate the performance of our method on real NLP tasks, in Table \ref{tab:glue}, we compare the performance of different models using the General Language Understanding Evaluation benchmark (GLUE) \cite{wang2018glue}. GLUE is a collection of resources for training, evaluating, and analyzing natural language understanding systems. Here we follow the BERT architecture \cite{devlin2018bert}. We pre-train all our models on wikitext-103 for 120k steps and then fine-tune on GLUE. The average score of our method is only 5 lower than the baseline but higher than other methods.

\begin{table*}[t]
    \caption{A comparison of the GLUE scores between the baseline (Softmax attention), kernel methods (Performer), sparse attention (Big Bird), and our method.}
    \centering
    \resizebox{0.95\linewidth}{!}{
        \begin{tabular}{@{} >{\bfseries}r|rrrrrrrr|r}
\toprule
\multicolumn{10}{c}{\textbf{\sffamily GLUE}} \\ \midrule
Model       &COLA (m)  &SST-2 (a)  &MRPC (f1/a)  &STS-B (p/s)  &QQP (f1/a)  &MNLI (a)  &QNLI (a)  &RTE (a)  &Average \\
\midrule
\midrule
Baseline    &48.7      &90.8       &89.2/84.6    &85.7/85.8    &85.9/89.8   &81.1      &87.9      &65.8     &81.3 \\ \midrule
Performer   &39.3      &90.1       &84.7/75.6    &81.0/80.7    &83.4/88.1   &76.6      &83.5      &61.3     &76.4 \\ \midrule
Big Bird    &30.2      &90.0       &83.3/78.7    &81.9/81.6    &83.5/87.5   &76.3      &83.3      &59.7     &76.1 \\ \midrule
Our         &40.1      &90.4       &84.6/75.9    &81.7/81.8    &83.7/88.2   &76.9      &83.4      &62.6     &\textbf{76.8} \\

\bottomrule
\end{tabular}
    }
    \label{tab:glue}
\end{table*}

\begin{table}[t]
    \caption{A comparison of perplexity between all the
    models on wikitext-103 and bookcorpus}
    \centering
    \resizebox{0.8\linewidth}{!}{
        \begin{tabular}{@{} >{\bfseries}r|r|r}
\toprule
Model          &wikitext-103 (ppl)  &bookcorpus (ppl)\\ 
\midrule
\midrule
Baseline       &5.665               &7.723\\ \midrule
Performer      &7.364               &8.957\\ \midrule
Big Bird       &7.798               &9.132\\ \midrule
GFormer        &6.837               &8.558\\
\bottomrule
\end{tabular}
    }
    \label{tab:perplexity}
\end{table}

\begin{figure}[h]
    \centering
    \includegraphics[width=1.0\linewidth]{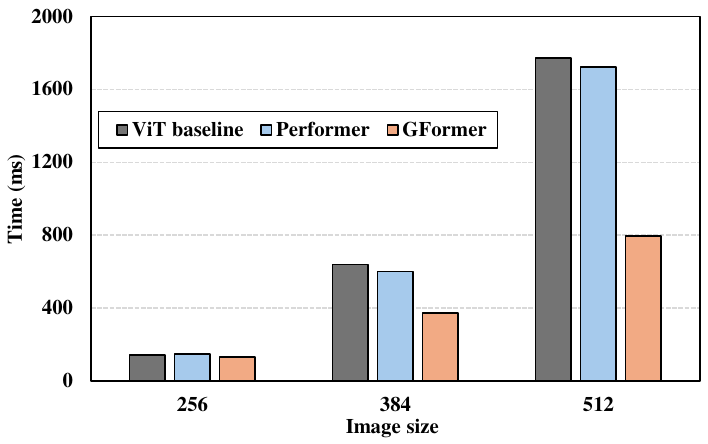}
    \vspace{-4mm}
    \caption{Performance of ViT with our attention on Gaudi.}
    \label{fig:vit-gaudi}
   \vspace{-2mm}
\end{figure}

\paragraph{\textbf{ViT Speedup}}
As shown in Figure \ref{fig:vit-gaudi}, the ViT model with Performer only achieves up to 1.1 $\times$ speedup over baseline. ViT model with GFormer achieves up to 2.2 $\times$ speedup over baseline since GFormer fully utilizes both MME and TPC. 

\paragraph{\textbf{ViT Accuracy}}
To evaluate different models' accuracy in Vision Transformer. We pre-train all models on ImageNet 2012 \cite{deng2009imagenet} (1 million images, 1,000 classes) at a resolution 384$\times$384. We set the patch size to 16$\times$16. As illustrated in Table \ref{tab:imagenet}, ViT with GFormer has only 1.4\% accuracy drop compared with baseline. 

\begin{table}[h]
    \caption{Top-1 Accuracy of pre-trained Vision Transformer base on ImageNet with different attention replacements. Acc. $\Delta$ represents the average accuracy drop to baseline.}
    \centering
    \resizebox{0.58\linewidth}{!}{
        \begin{tabular}{@{} >{\bfseries}rrr}
\toprule
Model          &Top-1 Acc  &Acc. $\Delta$ \\ 
\midrule
\midrule
Baseline       &81.5\%     &-     \\ \midrule
Performer      &79.9\%     &-1.6\% \\ \midrule
Big Bird       &79.6\%     &-1.9\% \\ \midrule 
GFormer        &80.1\%     &-1.4\%  \\
\bottomrule
\end{tabular}

    }
    \label{tab:imagenet}
\end{table}

\subsection{Comparison with GPUs.}
\begin{figure}[t]
    \centering
    \includegraphics[width=1.0\linewidth]{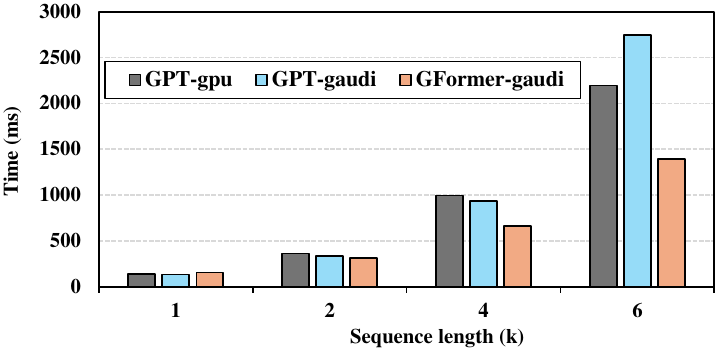}
    \vspace{-4mm}
    \caption{Performance of GPT on GPU and Gaudi.}
    \label{fig:gpt-gpu}
   \vspace{-2mm}
\end{figure}

We evaluate the performance of GPT and ViT on both GPU and Gaudi. The GPU type is V100 GPU with 32GB of memory on Bridges-2 \cite{brown2021bridges}. As shown in Figure \ref{fig:gpt-gpu}, we compare the GPT model with GFormer on Gaudi (GFormer-gaudi) with the GPT model with Softmax attention on Gaudi (GPT-gaudi) and the GPT model with Softmax attention on Gaudi on GPU (GPT-gpu). GPT-gaudi only has a similar performance to GPT-gpu. But it has worse performance when the sequence length reaches 6k. But our proposed GFormer-gaudi always has speedup over GPT-gpu, the speedup is up to 1.5 $\times$ when the sequence length is 6k.

\begin{figure}[t]
    \centering
    \includegraphics[width=1.0\linewidth]{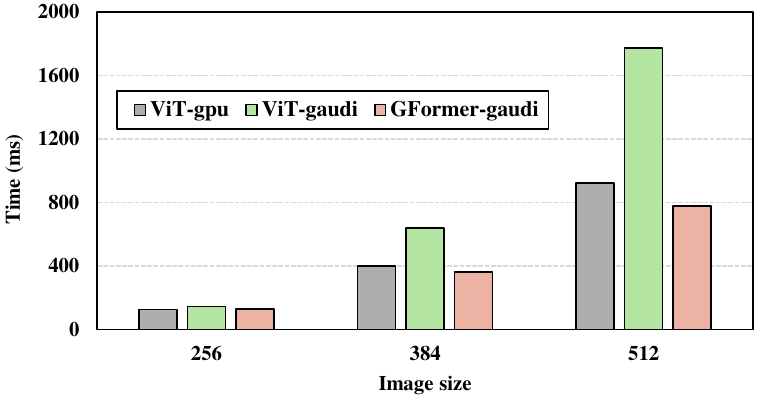}
    \vspace{-4mm}
    \caption{Performance of ViT on GPU and Gaudi.}
    \label{fig:vit-gpu}
   \vspace{-2mm}
\end{figure}

Figure \ref{fig:vit-gpu} shows the performance comparison among ViT with Softmax attention on GPU (ViT-gpu), ViT with Softmax attention on Gaudi (ViT-gaudi), and proposed ViT with GFormer on Gaudi (GFormer-gaudi). ViT-gaudi is 1.5 $\times$ on average slower than ViT-gpu. However, our proposed GFormer-gaudi achieves up to 1.2$\times$ speedup over Soft-gpu. These two experiments prove that transformer-based models on Gaudi can achieve speedup over GPU when we fully utilize its hardware resource.

\section{Related Work}
\label{sec:related}
Existing works exemplify the ongoing efforts to redesign attention mechanisms from both algorithm and algorithm-hardware co-deign aspects to balance computational efficiency with the powerful representational capabilities of Transformers. Specifically, Scatterbrain \cite{chen2021scatterbrain} presents a hybrid attention algorithm that combines the benefits of sparse and low-rank attention mechanisms. This approach aims to capture the advantages of both methods, offering a more flexible and efficient solution for approximating attention in Transformers. Even though this method showcases the higher attention approximation accuracy, it doesn't consider hardware efficiency.  

Works such as DOTA \cite{qu2022dota} and ViTALiTy \cite{dass2023vitality} exemplify specific hardware designs to meet the unique demands of Transformers. DOTA introduces a method to dynamically identify and omit attentions with minimal impact on performance, allowing for significant acceleration by reducing computational load. ViTALiTy, on the other hand, combines low-rank and sparse approximation techniques within a novel hardware design to accelerate Vision Transformers, showcasing the potential of custom hardware solutions to optimize performance efficiently.

On the other hand, several works leverage existing hardware to accelerate Transformer models effectively. Fang, et al. \cite{fang2022algorithm} presents a co-optimized framework that intelligently adjusts both the algorithmic and hardware aspects to exploit the sparsity of transformers, making it possible to achieve substantial acceleration without the need for specialized hardware. 
Yu, et al \cite{yu2023boost} discuss GPU-friendly 2:4 fine-grained structured sparsity and quantization for Transformers. 
Liu et al. \cite{liu2021transformer} exploit the dynamic sparsity in the attention of Transformers and provide corresponding GPU optimization. 
Zhao, et al. \cite{zhao2022fpga} introduce an FPGA-based Transformer accelerator with an output block stationary dataflow to minimize the repeated memory access by block-level and vector-level broadcasting. 

While specialized hardware designs offer increased flexibility, they necessitate knowledge of the hardware domain and pose challenges in real-world implementation. Our GFormer leverages existing hardware, enabling large-scale applications in real-world scenarios without additional effort. Unlike FPGA-based designs, which are complex to program, GFormer is user-friendly. Moreover, GFormer provides an economical alternative to GPU designs, delivering comparable functionality and performance at a lower cost. Targeting commercial, emerging heterogeneous hardware, \textit{GFormer underscores the potential of heterogeneous computing as an effective path for accelerating Transformer-based DL tasks}.

\section{Conclusion and Future Work}
\label{sec:conclusion}
Softmax operation is one of the major performance bottlenecks on the Gaudi processor, particularly when processing long sequences. In this work, we introduce an integrated approach that combines sparse and linear attention mechanisms. Our approach includes a windowed local-context self-attention TPC kernel and an efficient outer product TPC kernel for processing causal linear attention operations. We evaluate our proposed solution in GPT and ViT models on a Gaudi processor. The evaluation shows that our solution achieves up to 2 $\times$ and 2.2 $\times$ speedups in GPT and ViT compared to the original models using Softmax attention. 

In the future, we intend to initially extend our work to enable distributed LLM acceleration across multiple Gaudi cards, focusing on optimized communication. Subsequently, we will adapt our mixed attention mechanism for use with various heterogeneous AI accelerators, including Versal ACAP. Lastly, we aim to investigate the application of Gaudi in other emerging ML tasks, such as graph neural networks.



\clearpage
\bibliographystyle{ieeetr}
\bibliography{refs}
\end{document}